\documentclass[11pt]{article}
\usepackage{moriond,epsfig}

\def\be{\begin{equation}}
\def\ee{\end{equation}}
\def\bea{\begin{eqnarray}}
\def\eea{\end{eqnarray}}

\begin{document}
\vspace*{4cm}
\title{PHENOMENOLOGY OF SUSY WITH INTERMEDIATE SCALE PHYSICS}

\author{C. BIGGIO}

\address{Institut de F\'\i sica d'Altes Energies\\Universitat Aut\`onoma de Barcelona, 08193
 Bellaterra, Spain}

\maketitle\abstracts{The presence of fields at an intermediate scale
  between the Electroweak and the Grand Unification scale modifies the
  evolution of the gauge couplings and consequently the running of
  other parameters of the Minimal Supersymmetric Standard Model, such
  as gauginos and scalar masses. The net effect is a modification of
  the low energy spectrum which affects both the collider
  phenomenology and the dark matter relic density.}

\section{Introduction}

The presence of new physics at a scale intermediate between the
electroweak (EW) and the grand unified theory (GUT) scale is common to
many supersymmetric (SUSY) models. For example, in order to give mass
to the neutrinos, the minimal supersymmetric standard model (MSSM)
must be extended. The simplest extension consists in the addition of 2
or 3 chiral superfields which are singlets under the standard model
(SM) gauge group (type I seesaw~\cite{typeI}), but other extensions
are possible, such as the addition of a pair of ${\bf 15}+ {\bf
  \overline{15}}$ SU(5) representations (type II)~\cite{Rossi:2002zb}
or 1 (or 2 or 3) {\bf 24} (type
III)~\cite{Biggio:2010me,Esteves:2010ff}. Except for the singlet case,
the presence of new chiral superfields (and, in general, of new
fields) at an intermediate scale, affects the running of the MSSM
parameters, with a consequent distorsion of the low energy SUSY
spectrum.
Other examples are models where the breaking of the GUT symmetry to
the SM one is realized through intermediate steps or flavour models
with messengers. In the present work we restrict to consider only
chiral superfields at the intermediate scale: as we will see, their
main effect is to increase the value of the unified gauge
coupling. The presence of gauge fields would drive the result in the
opposite direction. However, as long as the net effect is the
enhancement of the unified coupling, the results shown here will be
qualitatively valid also in the presence of vector fields. In the
following we will describe the main effects on the running of the MSSM
parameters as well as some phenomenological consequences. We redirect
the readers to Ref.~\cite{BC2} for an extended analysis.

\section{MSSM running with intermediate scale}

In order to maintain gauge coupling unification, we
assume that only chiral superfields in complete vector-like
representations of SU(5) are present at the intermediate scale $M_I$. Even if the
unification is preserved, the running of gauge couplings is deflected,
above the scale $M_I$, by the presence of the new fields, as can be
observed in the left panel of Fig.~1. The net effect is an increment
of the unified gauge coupling, as can be seen by solving 1 loop RGEs:
\begin{equation}
 \frac{1}{\alpha_U} = \frac{1}{\alpha_i (M_Z)} 
- \frac{b^{SM}_i}{2 \pi} \ln \frac{M_S}{M_Z} 
- \frac{b^0_i}{2 \pi} \ln \frac{M_{\rm GUT}}{M_S} 
- \frac{\Delta b}{2 \pi} \ln \frac{M_{\rm GUT}}{M_I} \equiv \frac{1}{\alpha^0_U} 
- \frac{\Delta b}{2 \pi} \ln \frac{M_{\rm GUT}}{M_I} \,, 
\label{eq:alpharun}
\end{equation}
where $\alpha^0_U$ is the unified coupling in the MSSM without
intermediate scale ($\alpha^0_U\simeq 1/25$), $b_i^{SM} =
(41/10,-19/6,-7)$ and $b^0_i = (33/5,1,-3)$ are respectively the SM
and MSSM $\beta$-function coefficients for $\alpha_i$ ($i=1,2,3$),
$\Delta b$ is the universal contribution of the additional fields at
$M_I$, given by the sum of the Dynkin indexes of the SU(5)
representations, and $M_S$ is the typical low-energy SUSY scale.  From
Eq.~(\ref{eq:alpharun}), we see that, since $\Delta b \geq 0$ for
chiral superfields, the unified coupling $\alpha_U$ is in general
larger than the MSSM one.

The increment of $\alpha_U$ is the principal effect of the presence of
the intermediate scale and the one which drives all the
others. Moreover, by requiring the perturbativity of $\alpha_U$ up to
the GUT scale, a large portion of the parameter space can already be
excluded, as it is shown in the right panel of Fig.~1, where the white
area corresponds to a Landau pole below $M_{GUT}$. Notice that we are
considering 1 loop RGEs. We have checked~\cite{BC2} that the
consequence of two loops running is to strengthen the effect
especially close to the Landau pole, while it is almost irrelevant for
low $\Delta b$ and/or large $M_I$. The perturbativity bounds are then
slightly stronger than the ones shown here and the allowed region is
consequently smaller (also for the other plots in the following).
\begin{figure}
\begin{center}
\psfig{figure=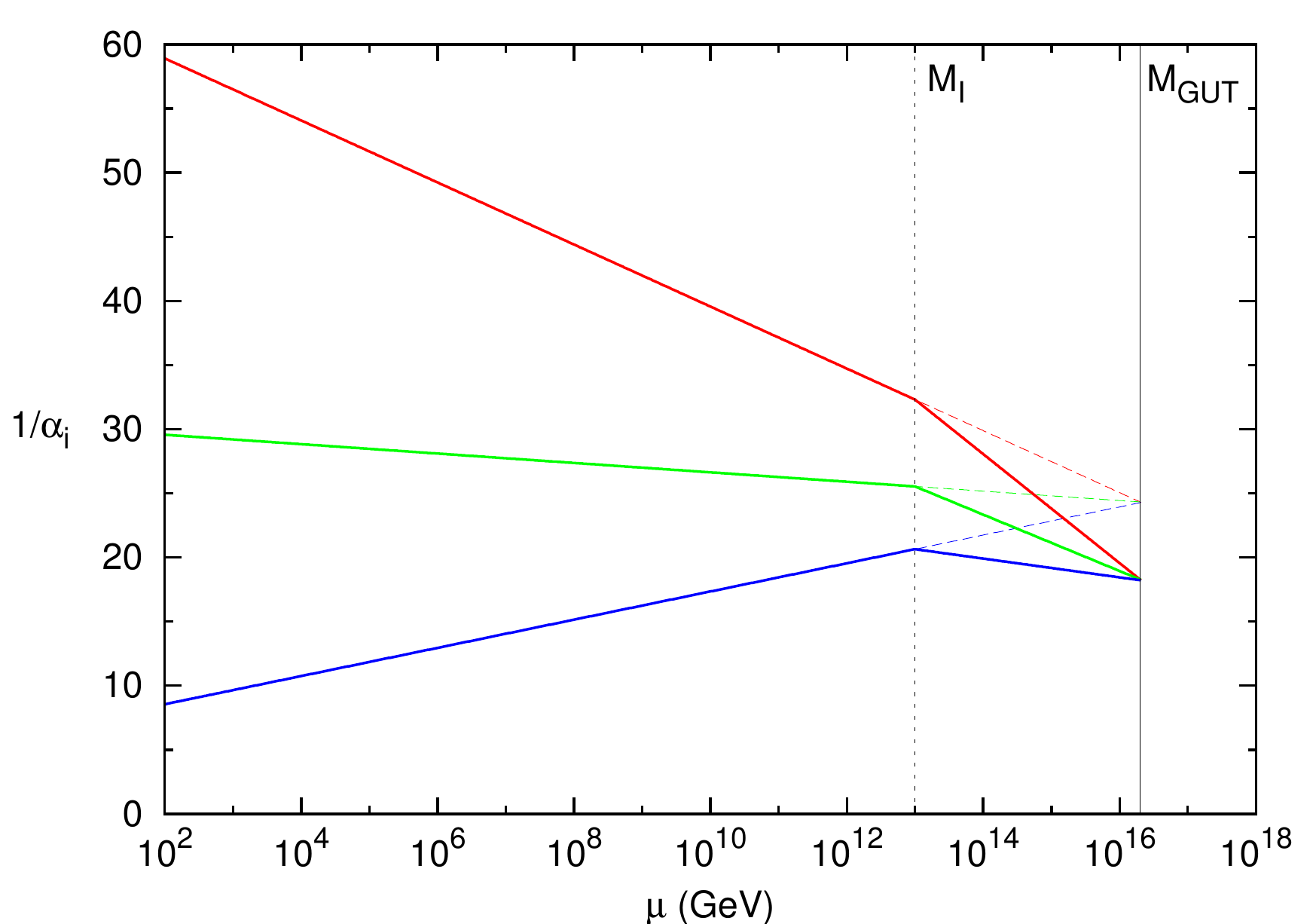,width=0.4\textwidth}
\hspace{1cm}
\psfig{figure=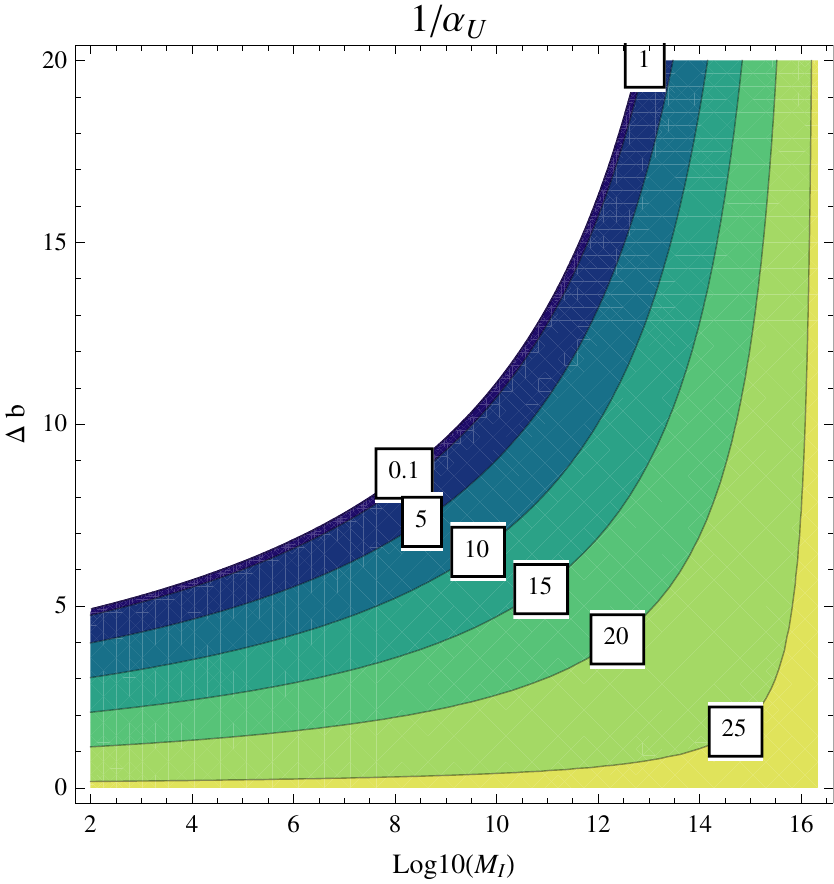,width=0.3\textwidth}
\end{center}
\caption{Left: modification of the gauge couplings running in
  the presence of matter at the intermediate scale $M_I = 10^{13}$ GeV
  with $\Delta b = 5$ (corresponding, for instance, to a copy of {\bf
    24}); the dashed lines correspond to the MSSM
  running. Right: contours on the plane $M_I$-$\Delta b$ of the
  inverse of the unified gauge coupling $1/\alpha_U$.}
\label{fig:alpha}
\end{figure}

Let us now move to consider the effect of the intermediate scale on
the other MSSM parameters.  Since the RGEs of gaugino masses $M_i$ are
strictly related to the ones of the gauge couplings, the increase of
$\alpha_i$ above $M_I$ will determine a faster running of each
$M_i$. However, since we are assuming gaugino mass unification, the
MSSM low energy ratio of gaugino masses will be maintained and the same
MSSM low energy spectrum could be obtained by a simple rescaling of
the unified gaugino mass $M_{1/2}$, which would become larger. Then,
if only gauginos were there, no interesting observable effect would be
produced.

\begin{figure}[!t]
\begin{center}
\psfig{figure=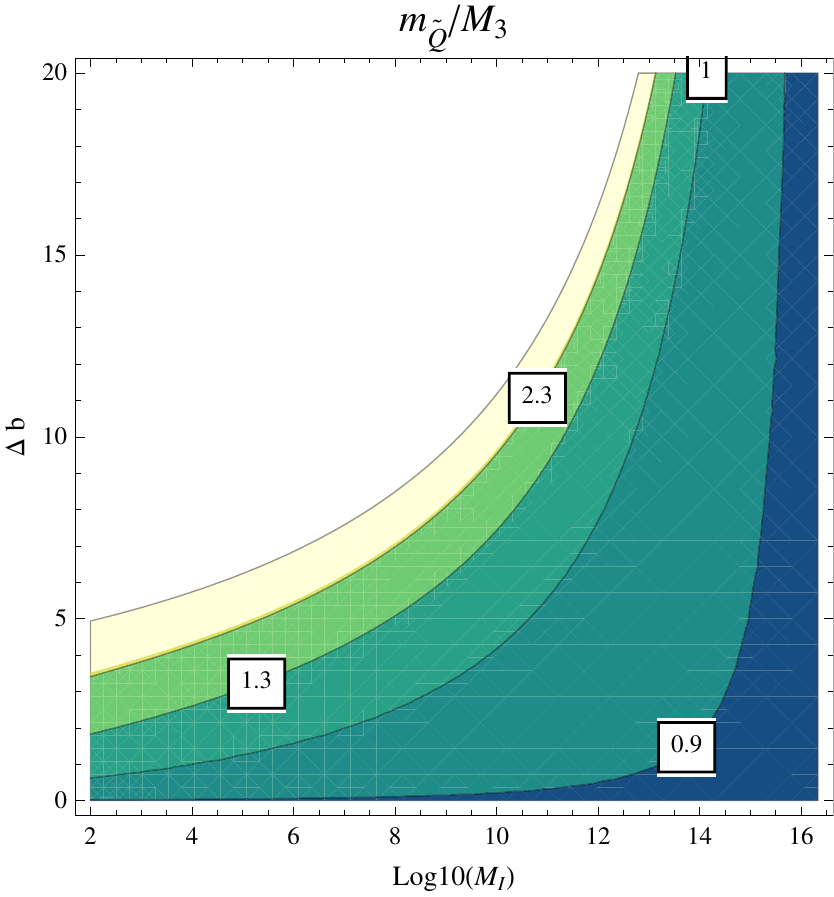,width=0.3\textwidth}
\hspace{.2cm}
\psfig{figure=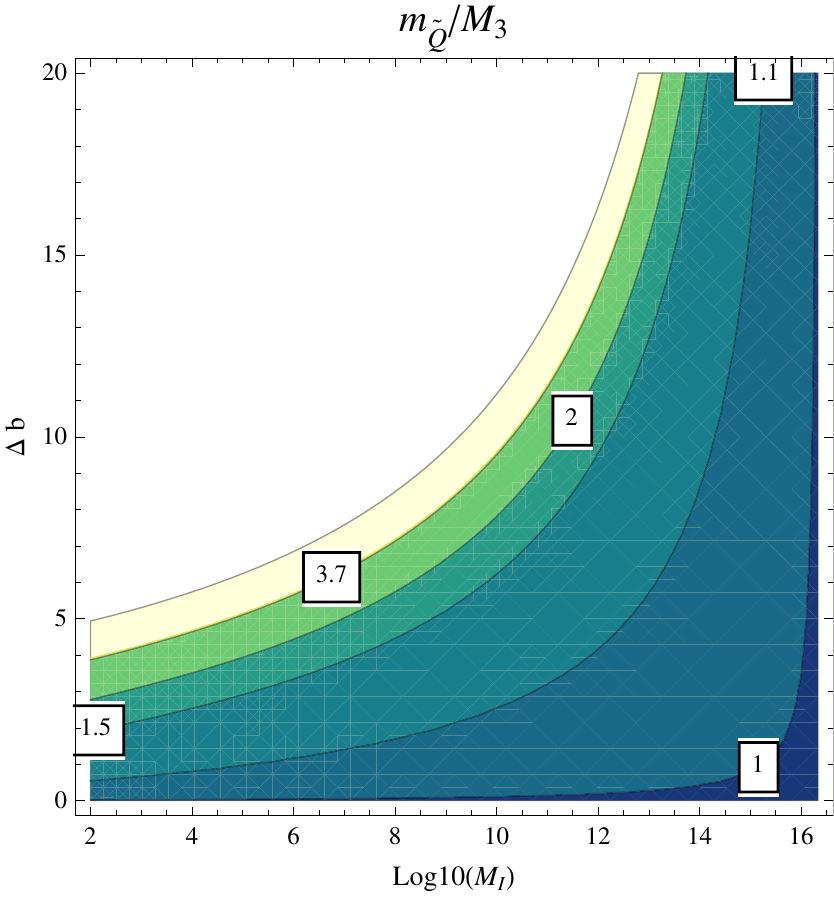,width=0.3\textwidth}
\hspace{.2cm}
\psfig{figure=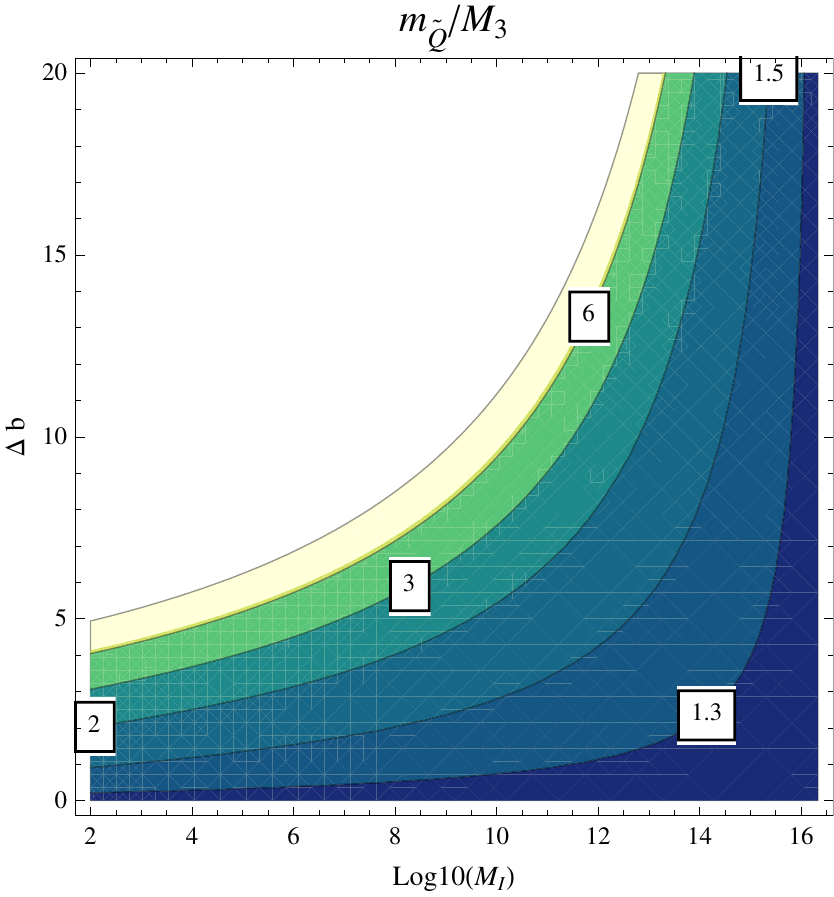,width=0.3\textwidth}
\end{center}
\caption{Ratio of the 1st or 2nd generation LH squarks over the gluino
  mass $M_3$ for $m_{\tilde Q}/M_{3}=0,\, 1,\, 2$ at $M_{\rm GUT}$.}
\label{fig:ratio}
\end{figure}

On the other hand, if we consider scalar masses, we observe a
different situation: the increment of gauge couplings and gaugino
masses above $M_I$ will determine an increase of the low energy scalar
masses with respect to the MSSM case, for the same low energy gaugino
mass spectra. This is because, to obtain the same low energy gaugino
masses, a higher value for $M_{1/2}$ is needed which, together with
the larger values of $\alpha_i$ above $M_I$, enhances the gauge part
of the running of scalar masses in the upper part of the RG flow,
generating larger values for low energy masses.  In a nutshell, the
net effect of the intermediate scale is to increase the ratio of
scalar over gaugino masses. This is shown in Fig.~\ref{fig:ratio},
where the ratio of the first generations left-handed squark over
the gluino mass is plotted.

This has interesting phenomenological consequences, both for what
concerns SUSY searches at the LHC and for the dark matter (DM)
phenomenology.

\section{LHC Observables}

By looking at the ratio of the first generation left-handed squark
over gluino masses in Fig.~\ref{fig:ratio}, we already see a potential
consequence of the presence of intermediate scale for SUSY collider
searches: the configuration $M_3\approx m_{\tilde{Q}}$, that gives the
highest sensitivity at the LHC, could be never reached, since the
tendency is to have heavier squarks. However, if this will be the
case, no definite conclusion will be extract, since a similar
behaviour could be obtained simply by increasing the value of the high
energy scalar mass (see the right panels of Fig.~\ref{fig:ratio}). On
the other hand, if the case $M_3\approx m_{\tilde{Q}}$ will be
observed, this would give an upper bound on $\alpha_U$ and strongly
constrain the intermediate scale parameters.

In order to obtain clear information on the intermediate scale
physics, observables as much indipendent as possible of the high
energy parameters are necessary. If we assume gaugino mass unification
and consider one loop RGEs, the gaugino and first generations sfermion
masses can be written as:
\begin{eqnarray}
M_i(M_S) &=& A_i(M_S,\Delta b,M_I)\, M_{1/2} \\
m^ 2_{\tilde{f}}(M_S) &=& m_{\tilde f}^2(M_{\rm GUT}) + B_{\tilde{f}}(M_S,\Delta b,M_I)\, M_{1/2}^2 \,, 
\end{eqnarray}
where the coefficients $A_i$ and $B_{\tilde{f}}$ are functions of
$\Delta b$, $M_I$ and $M_S$.  It is then clear that in the
combination
\begin{equation}
\Delta^{ff^\prime}_i\equiv \frac{m^ 2_{\tilde{f}}-m^ 2_{\tilde{f'}}}{M_i^2}
\label{eq:massinv}
\end{equation}
the explicit dependence on the GUT-scale parameters drops, if
$m_{\tilde f^\prime}^2(M_{\rm GUT})=m_{\tilde f}^2(M_{\rm GUT})\equiv
m_0^2$ as in CMSSM-like scenarios or, more in general, in the case of
GUT-symmetric initial conditions (a well-motivated assumption in our
setup, as we are requiring unification).

\begin{figure}[!t]
\begin{center}
\psfig{figure=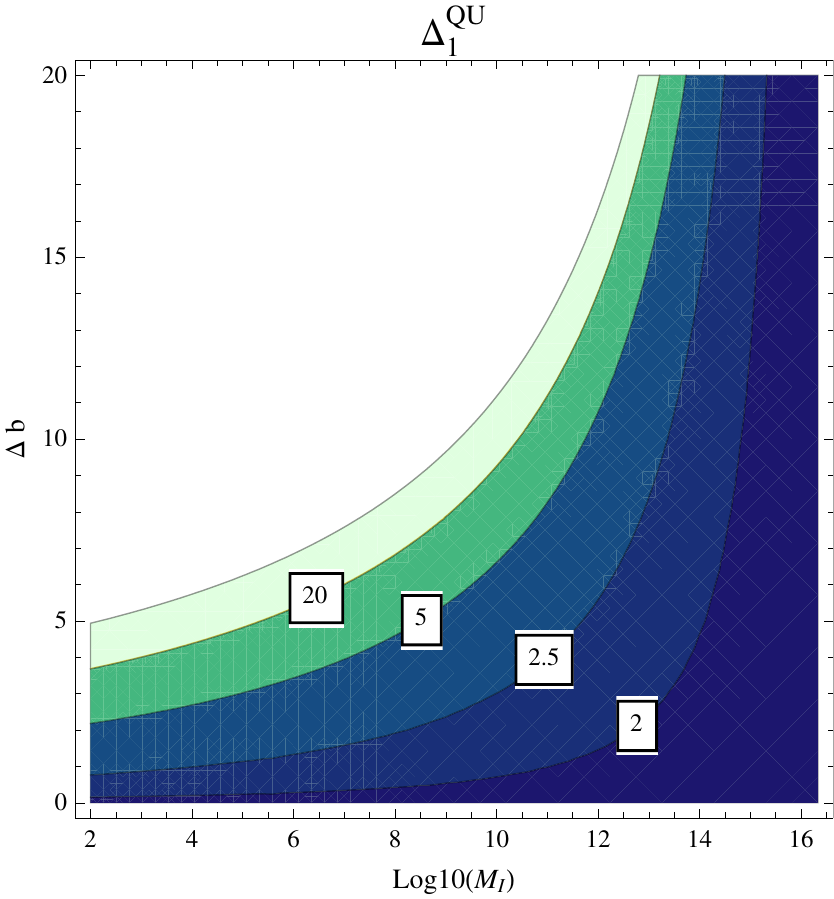,width=0.3\textwidth}
\hspace{.2cm}
\psfig{figure=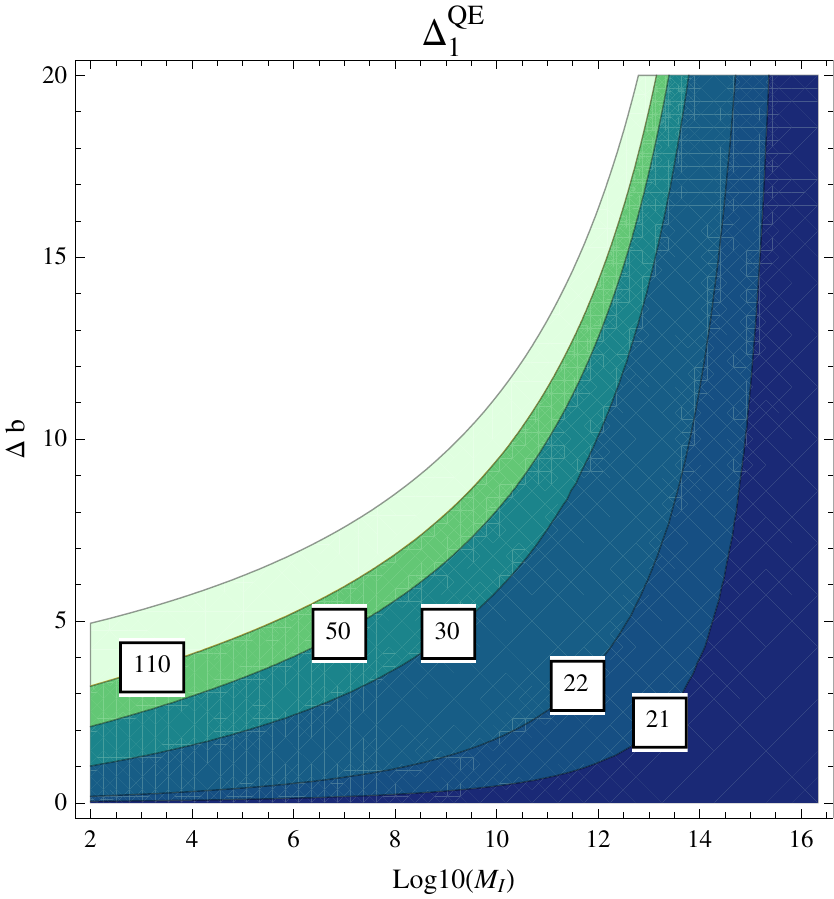,width=0.3\textwidth}
\hspace{.2cm}
\psfig{figure=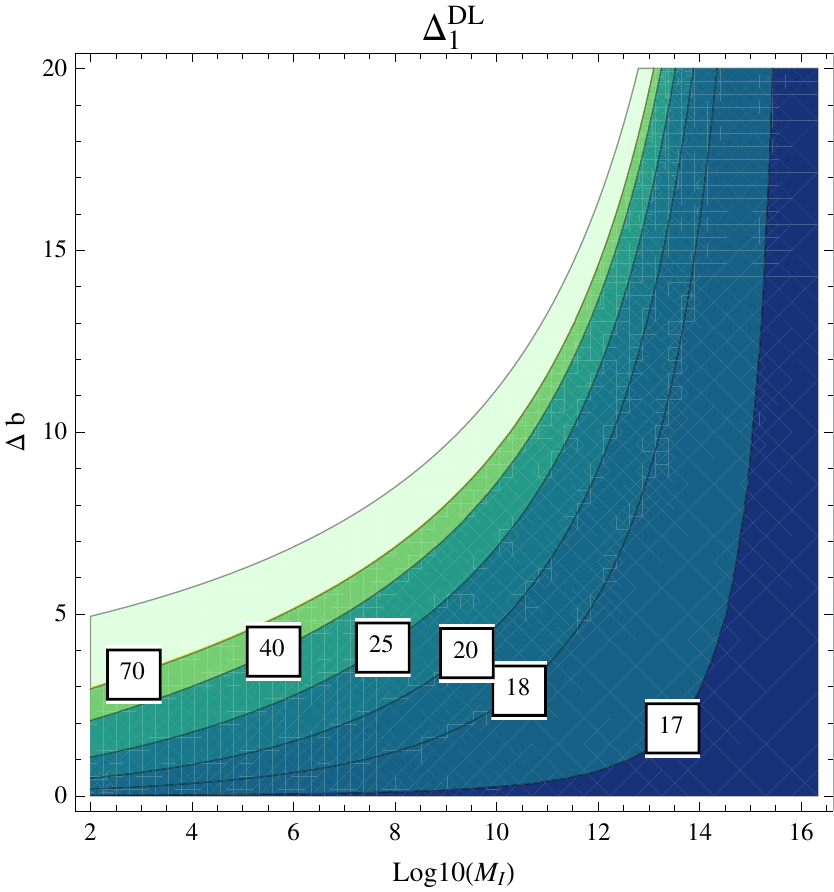,width=0.3\textwidth}
\end{center}
\caption{Contours for the mass invariants $\Delta^{QU}_1$,
  $\Delta^{QE}_1$, $\Delta^{DL}_1$ for M$_{S}=1$TeV.}
\label{fig:inv}
\end{figure}

As in Ref.~\cite{Buckley:2006nv}, where it has been suggested to use
these invariants to discriminate among different SUSY seesaw models,
we consider the SU(5)-inspired combinations:
\begin{equation}
\Delta^{QU}_1 \equiv\frac{m^2_{\tilde{Q}}-m^2_{\tilde{U}}}{M_1^2},\quad\quad \Delta^{QE}_1 \equiv\frac{m^2_{\tilde{Q}}-m^2_{\tilde{E}}}{M_1^2},
\quad\quad \Delta^{DL}_1 \equiv\frac{m^2_{\tilde{D}}-m^ 2_{\tilde{L}}}{M_1^2}.
\label{eq:SU5inv}
\end{equation}
Contours for these quantities on the $(M_I, \Delta b)$ plane are shown in
Fig.~\ref{fig:inv} (taking $M_S = 1$ TeV).  As we can see, the
invariants rapidly grow for increasing $\alpha_U$. If the SUSY
spectrum will be measured at the LHC with enough precision, from these
invariants it should be possibile to test if the spectrum is MSSM-like
or not, and, in the latter case, even to distinguish between the
presence of intermediate scale physics or non-universal boundary
conditions, if correlation among these invariants are
verified.\footnote{In Ref.~\cite{BC2} also another set of invariants
  is considered, namely ratios of differences of scalar masses. These
  have the advantage of being less dependent on the scale
  $M_S$. Analogous plots can be obtained.}

\begin{figure}
\begin{center}
\psfig{figure=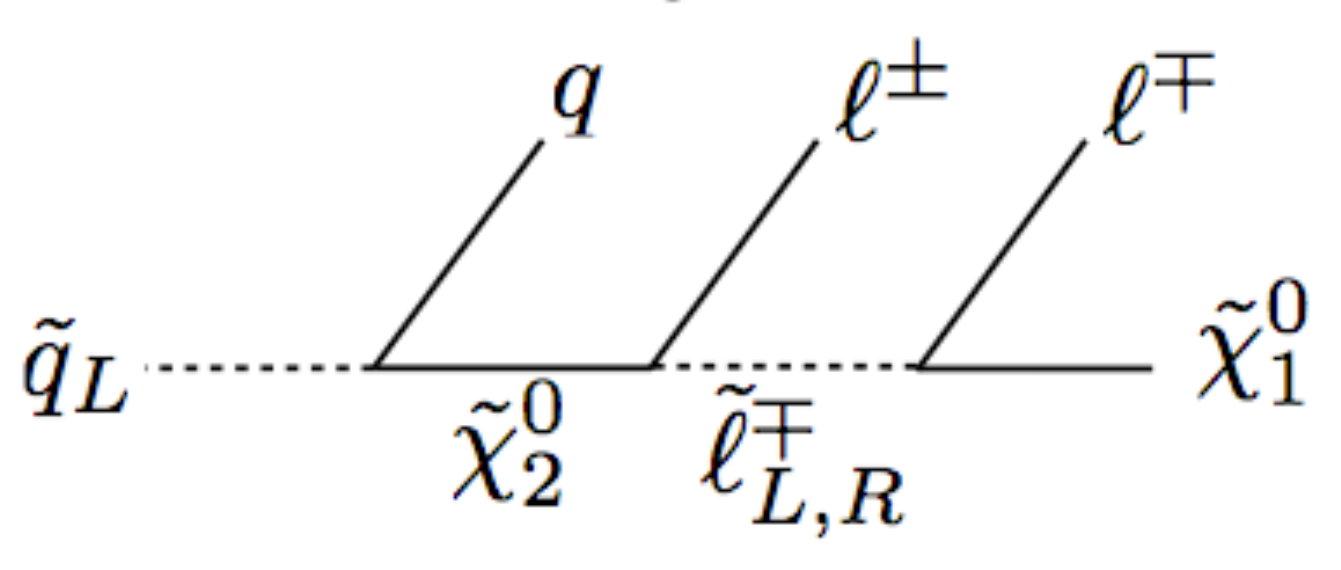,width=0.4\textwidth}
\hspace{1cm}
\psfig{figure=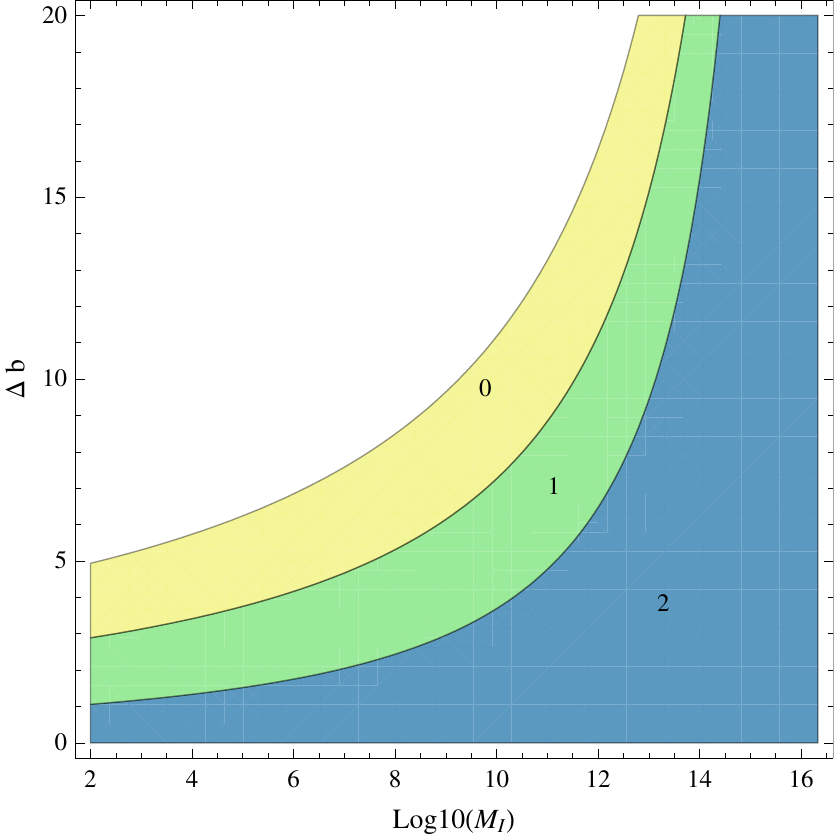,width=0.3\textwidth}
\end{center}
\caption{Left panel: an example of cascade decay. Right panel: maximum
  number of edges.}
\label{fig:edges}
\end{figure}

A possible way to measure sparticle masses at the LHC is through
cascade decays like the one depicted in Fig.~\ref{fig:edges}. The
intermediate particles are real if
\begin{equation}
m_{\tilde{Q}} > m_{{\tilde\chi}^0_2} > m_{\tilde{\ell}_{L,R}} > m_{{\tilde\chi}^0_1} \, .
\label{eq:cascade}
\end{equation}
In this case the invariant mass distributions of the outgoing SM
particles (jets and isolated leptons) exihibit sharp kinematic
end-points. By measuring the position of these endpoints in different
invariant mass distributions, the masses of the involved sparticles
can be measured~\cite{Bachacou:1999zb}. The presence of intermediate
scale physics moves the position of the edges and, if independent
measurements of sparticle masses and edges were available, then
important information on the intermediate scale could be extracted.

Moreover, notice that, depending on the spectrum, zero, one or two sharp edges
can be present. Indeed, this depends if the above condition is
satisfied by both $m_{\tilde{\ell}_L}$ and $m_{\tilde{\ell}_R}$ or by
one of them (typically $m_{\tilde{\ell}_R}$) or none. Since the effect
of the intermediate scale is precisely that of enhancing the ratio of
scalar over gaugino masses, the above inequality will be more hardly
satisfied in the presence of intermediate scale physics. In the right
panel of Fig.~\ref{fig:edges} the maximum number of edges on the plane
$(M_I,\Delta b)$ is depicted. It is then clear that the observation of
two or one edge in the mass invariant distributions would rule out a
large portion of parameter space.

\section{Neutralino Dark Matter}
\begin{figure}[!t]
\begin{center}
\psfig{figure=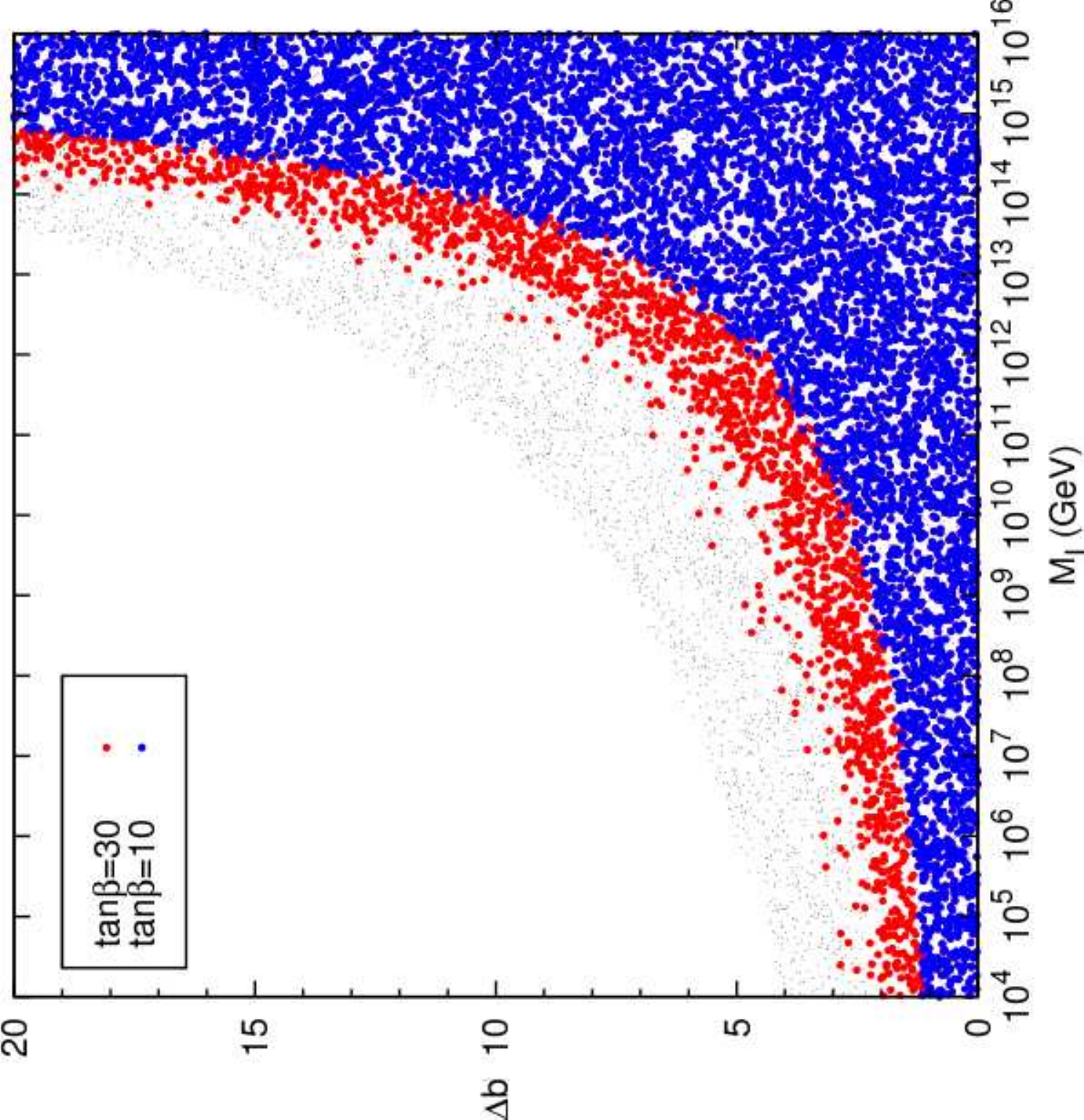,width=0.3\textwidth,angle=-90}
\hspace{1cm}
\psfig{figure=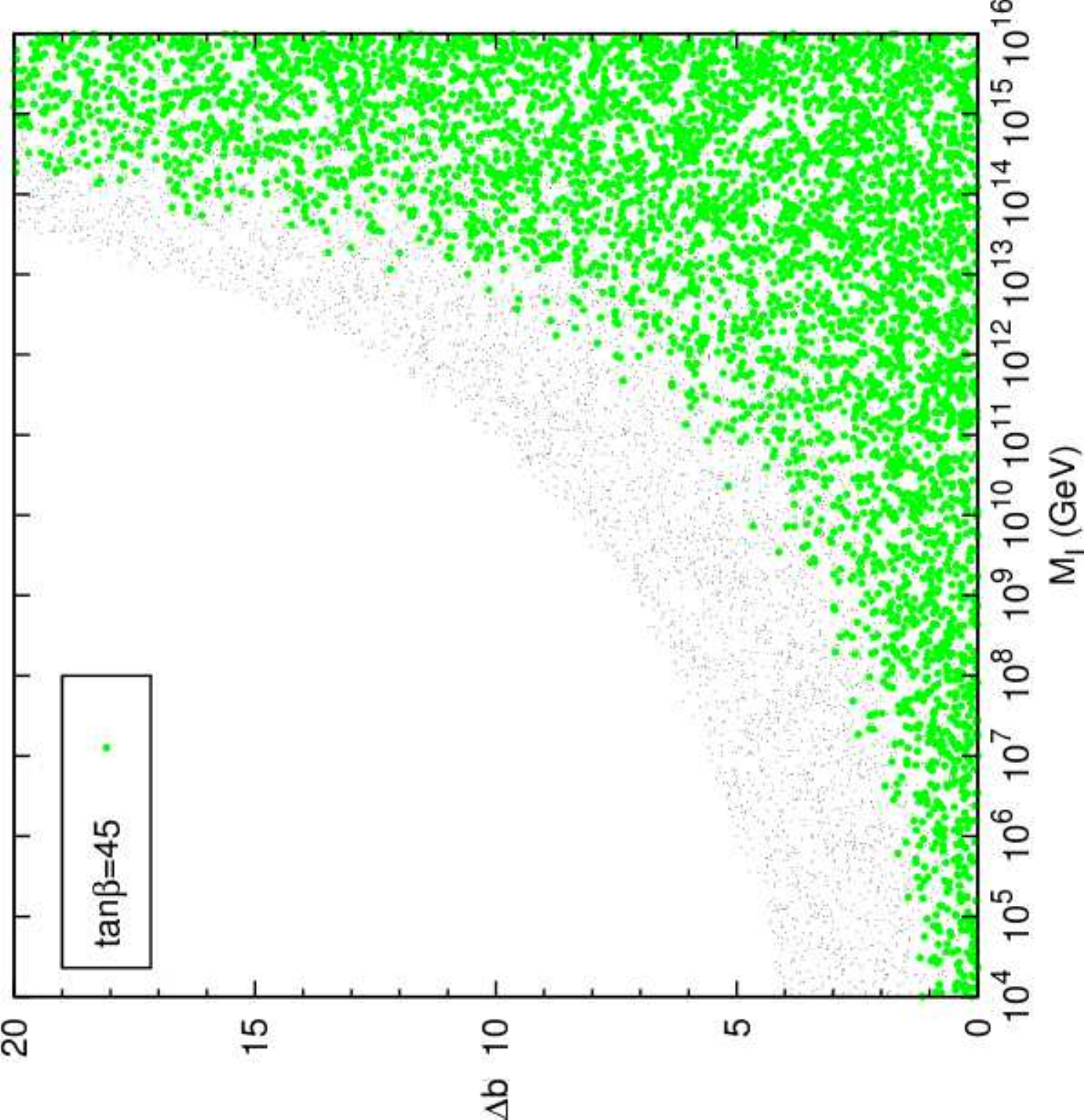,width=0.3\textwidth,angle=-90}
\end{center}
\caption{Left: region on the $(M_I, \Delta b)$ plane where the correct
  relic density for $\tilde{\chi}^0_1$ is obtained via
  $\tilde{\tau}$ coannihilation (blue points corresponds to $\tan
  \beta=10$, red to $\tan \beta=30$). Right: the same
  for the A-funnel region.}
\label{fig:DM}
\end{figure}

The modification of the SUSY spectrum described in the previous
sections can destabilise the regions of the parameter space where
precise relations among the parameters are required in order to
fulfill the WMAP bound on DM relic abundance. Indeed the lightest
neutralino $\chi^0_1$ is overproduced in the early universe and the
measured DM relic density is not obtained unless the neutralino
(co)annihilation cross section is enhanced by particular conditions.
Such conditions define few regions in the parameter space where the
WMAP bound is satisfied: (i) the $\tilde{\tau}$ coannihilation region,
where the correct relic density is achieved thanks to an efficient
$\tilde{\tau}$-$\tilde{\chi}^0_1$ coannihilation, which requires
$m_{\tilde{\tau}_1} \approx m_{\tilde{\chi}^0_1}$; (ii) the
``focus-point'' region, where the Higgsino-component of
$\tilde{\chi}^0_1$ is sizable, i.e.~$\mu \approx M_1$; (iii) the
A-funnel region, where the neutralino annihilation is enhanced by a
resonant s-channel CP-odd Higgs exchange, if $m_{A} \simeq 2\times
m_{\tilde{\chi}^0_1}$. Notice that typically the neutralino is
$\tilde{B}$-like, which means $m_{\chi^0_1}\approx M_1$. Then it is
clear that, since the intermediate scale tends to increase the ratio
of scalar over gaugino masses, the high-energy parameter space regions
where the above conditions are satisfied will be distorted and could
eventually disappear. 

We first consider the $\tilde{\tau}$ coannihilation region. In the
constrained MSSM it usually corresponds to a thin strip on the border
of an area excluded because it gives a $\tilde{\tau}$ lightest SUSY
particle. It has already been noticed in specific
models~\cite{Calibbi:2007bk,Esteves:2009qr,Biggio:2010me,Esteves:2010ff,Esteves:2011gk}
that the introduction of new physics at intermediate scale distorts
this region. If the effect is large enough the $\tilde{\tau}$ will
never be ligher than $\chi^0_1$ and the coannihilation region will
disappear. In the left panel of Fig.~\ref{fig:DM} the blue points
represent the area on the $(M_I,\Delta b)$ plane where the
coannihilation is possible for $\tan \beta=10$, while the red ones
correspond to $\tan \beta=30$ (grey dots correspond to the region allowed by the perturbativity bounds).\footnote{The results of this section
  have been obtained performing numerical two loops computations.}
Even if the allowed area can be enlarged for even higher $\tan\beta$,
for $\alpha_U \geq 0.1$ the coannihilation region will anyway
disappear.

A similar fate will follow the A-funnel region, as it is shown in the
right panel of Fig.~\ref{fig:DM} for $\tan\beta = 45$. Indeed in this
case the mass of the CP-odd scalar Higgs is given by $ m^2_A \approx
m^2_{H_d} - m^2_{H_u}$ and, since the ratios $|m^2_{H_{u,d}}|/M_1$
grow with $\alpha_U$, $m_A/M_1$ gets increased too. Again, for large
enough $\alpha_U$, the correct neutralino relic density cannot be
obtained with such a mechanism.

The situation with the focus point region is a bit different. Even if
the intermediate scale tends generically to increase the ratio
$\mu/M_1$ rendering the neutralino more and more $\tilde{B}$-like, it
is always possible to find configurations with $\mu \approx M_1$, such
that the Higgsino component of $\tilde{\chi}^0_1$ is sufficiently
large to give a sizeable annihilation cross-section.  The focus point
region is therefore the only DM branch which is not destabilised by
the intermediate scale, if CMSSM-like boundary conditions are assumed.

\section{Conclusions}

In this talk we have discussed the main consequences of the presence
of chiral superfields at a scale intermediate between the EW and the
GUT scale. The main effect is the increment of the unified gauge
coupling, which causes the increment of the ratio of scalar over
gaugino masses. This has interesting phenomenological consequences
that we have discussed here. In particular it destabilises the regions
of the parameter space where the correct DM relic density can be
achieved; only the focus-point region is practically unaffected. The
presence of such new physics can be tested indirectly at the LHC by
looking at the (number of) edges in cascade decays and/or by building
mass invariants, once the sparticle spectrum is measured. Therefore we
have shown that precious information, i.e. mainly constraints, on
intermediate scale physics can be obtained by considering the
appropriate low energy observables.

\section*{Acknowledgments}

I thank the organisers of this conference for the pleasant and
exciting atmosphere.


\end{document}